\documentstyle[12pt]{article}

\makeatletter

\@addtoreset{equation}{section}
\makeatother

\begin{document}
\begin{flushright} 
DPNU-96-63 \\ 
December 1996 
\end{flushright} 

\vspace{2cm}

\begin{center}
{\huge Effective Messenger Sector from \\
\vspace{2mm}
Dynamical Supersymmetry Breaking} 
\end{center}

\vspace{1cm}

\begin{center}
{\large Naoyuki Haba, Nobuhito Maru and Takeo Matsuoka} 
\end{center}

\begin{center}
{\it Department of Physics \\ 
Nagoya University \\ 
Nagoya 464-01, Japan} 
\end{center}

\begin{center}
{\normalsize {\tt haba@eken.phys.nagoya-u.ac.jp}} \\ 
{\normalsize {\tt maru@eken.phys.nagoya-u.ac.jp}} \\ 
{\normalsize {\tt matsuoka@eken.phys.nagoya-u.ac.jp}} 
\end{center}

\vspace{3cm}
\setcounter{page}{0}
\thispagestyle{empty}
\begin{center}
{\bf Abstract}
\end{center}

\begin{center}
\begin{minipage}[t]{13cm}
In the framework of the dynamical 
supersymmetry breaking 
we construct the messenger sector as 
the effective theory of supersymmetry breaking sector, 
which is based on $SU(3) \times SU(2)$ model of 
Affleck, Dine and Seiberg. 
In our model, messenger superfields with 
the non-renormalizable interaction are contained. 
By minimizing the scalar potential, we show that 
the supersymmetry breaking can be communicated to 
the visible sector without breaking QCD color. 
In this model there appear various scales. 
Supersymmetry breaking scale turns out to be 
the intermediate scale ($\sim 10^{10}$ GeV) 
between the GUT scale and the soft supersymmetry 
breaking scale. 
\end{minipage}
\end{center}
\newpage
\baselineskip 8mm
\section{Introduction} 
\indent

In order for supersymmetry to be relevant to 
the real world, it must be broken. 
Dynamical supersymmetry breaking \cite{Witten82} 
is one of the attractive scenarios 
because it can solve 
the gauge hierarchy problem elegantly 
by the dimensional transmutation. From the phenomenological 
point of view, it is a problem 
of how supersymmetry breaking is communicated to the 
visible sector, in other words, 
what is the messenger interaction. 
At present, we know of two scenarios in this regard. 
One is a usual hidden sector scenario \cite{Nilles}, 
in which the gravity plays the role of the messenger 
of supersymmetry breaking. 
Another is a gauge-mediated supersymmetry breaking 
scenario \cite{GMSB,DN}, in which the gauge interactions 
play the role of the messenger of supersymmetry breaking. 
One of the attractive advantages of the latter scenario 
is that the flavor changing neutral currents (FCNC) are 
naturally suppressed, 
because gauginos, squarks, sleptons masses appear 
radiatively at the low energy, contrary to 
the hidden sector scenario and are proportional to the 
flavor-blind gauge coupling constants squared. 
Furthermore, this scenario is highly predictive 
because the superparticle spectrum is calculable.

Recently, Dasgupta, Dobrescu and Randall \cite{Randall} 
showed that the true vacuum does not preserve 
QCD color in the minimal gauge-mediated 
supersymmetry breaking scenario. 
In the minimal gauge-mediated 
supersymmetry breaking \cite{DN}, 
the messenger superpotential is of the form 
\begin{equation}
W_{{\rm mes}} = k_1 \phi^+ \phi^- X + 
\frac{1}{3} \lambda X^3 
+ k_3 X l\bar{l} + k_4 X q\bar{q}, 
\end{equation}
where $q, \bar{q}$ are messenger quarks 
in {\bf 3} and ${\bf \bar{3}}$ of 
the color $SU(3)_C$, $l, \bar{l}$ 
are messenger leptons in {\bf 2} 
of $SU(2)_L$, $X$ 
is a singlet and $\phi^+,  \phi^-$ 
have $+1, -1$ charges of the messenger group $U(1)_m$. 
In this conventional model the $\phi^+  \phi^- X$ term 
communicates supersymmetry breaking from 
the charged fields under $U(1)_m$ to 
the singlet $X$ and the $X q \bar{q}$ 
and $X l\bar{l}$ terms link the singlet with 
the messenger quarks and leptons. From Eq. (1.1) 
the scalar potential of the messenger sector is given by 
\begin{eqnarray}
V_{{\rm mes}} &=& k_1^2 |X|^2 (|\phi^+|^2 + |\phi^-|^2) 
+ |k_1 \phi^+ \phi^- + \lambda X^2 
+ k_3 l\bar{l} + k_4 q\bar{q}|^2 \nonumber \\ 
&& + k_3^2 |X|^2 (|l|^2 + |\bar{l}|^2) + 
k_4^2 |X|^2 (|q|^2 + |\bar{q}|^2) \\ 
&& + M'^2 (|\phi^+|^2 + |\phi^-|^2) + 
\frac{g^2_1}{2} (|\phi^+|^2 - |\phi^-|^2)^2 \nonumber, 
\end{eqnarray}
where the second last term represents the soft supersymmetry 
breaking term with the mass $M'^2 < 0$ and 
the last term is $U(1)_m$ D-term. 
Since the soft supersymmetry breaking mass is negative, 
the vacuum expectation values (VEVs) of $\phi^+, \phi^-$ 
become non-zero. 
Then the supersymmetry breaking is transmitted to 
the singlet through the first term in Eq. (1.1). 
Through the minimization of Eq. (1.2), 
$X$ and $F_X$ are expected to develop 
non-zero VEVs. If this is the case, 
the supersymmetry breaking is transmitted to 
the visible sector radiatively 
by way of the last two terms in Eq. (1.1). 
As pointed out in Ref. \cite{Randall}, however, 
careful study shows that QCD color is violated 
in the true vacuum.  We are now at the stage that 
we have to reconsider the gauge-mediated 
supersymmetry breaking scenario and extend 
the messenger sector and/or the messenger interactions. 
Many attempts \cite{Many} have been made on 
the variations of the 
minimal gauge-mediated supersymmetry breaking.

In this paper we construct the messenger sector 
as the effective theory of supersymmetry 
breaking sector, 
which is based on $SU(3) \times SU(2)$ 
model of Affleck, Dine, 
and Seiberg \cite{ADS85}. 
In the conventional model of the gauge-mediated 
supersymmetry breaking 
all of the messenger matter fields and interactions 
are put by hand. 
The messenger superpotential (1.1) does not incorporate 
the F-type of supersymmetry breaking. 
The supersymmetry breaking is put into the messenger 
sector only through the soft 
scalar masses of $\phi^+$ and $\phi^-$. 
In contrast with the conventional model, 
our messenger sector superpotential 
is the effective superpotential of the supersymmetry 
breaking sector and no superfields are added to 
the messenger sector superpotential. 
This messenger superpotential yields the F-type of 
supersymmetry breaking. 
What we would like to emphasize here is that 
it does not seem to be appropriate to analyze 
the supersymmetry breaking sector and 
the messenger sector separately. From the beginning 
the messenger superfields with the non-renormalizable 
interaction are contained in the 
supersymmetry breaking sector. 
Furthermore, we show that supersymmetry breaking can be 
communicated to the visible sector 
without breaking QCD color.

As discussed in Ref. \cite{DN}, 
gaugino masses are generated at one-loop, 
whereas squark and slepton masses are generated 
at two-loop. As a result, FCNC are naturally suppressed. 
We also take the effects of gravitational 
interaction into account 
in calculating masses of gauginos, squarks and sleptons. 
It is found that 
these effects are not important compared with 
gauge interaction effects. 
In the present framework, we find that 
supersymmetry is broken at the intermediate 
scale ($\sim 10^{10}$ GeV) between the GUT scale and 
the soft supersymmetry breaking scale. 
The masses of the messenger fields ($10^{6 \sim 9}$ GeV) are 
in the intermediate scale 
between supersymmetry breaking scale and soft supersymmetry 
breaking scale.

This paper is organized as follows. In Section 2, the dynamical 
supersymmetry breaking sector is discussed. 
We construct the messenger sector as the effective 
theory of dynamical supersymmetry breaking sector 
and minimize the scalar potential. 
In Section 3, we estimate various scales in our model. 
Summary is found in Section 4.

\section{Messenger sector as the effective theory 
of supersymmetry breaking sector}
\indent

Our model is based on the $SU(3) \times SU(2)$ model 
\cite{ADS85}, which breaks supersymmetry dynamically. 
This model has the $SU(3) \times SU(2)$ 
gauge group, a global $U(1)_m$ and 
a non-anomalous global R-symmetry. 
In our model this $U(1)_m$ is gauged from the beginning. 
The representations and charges of the matter fields 
are summarized in Table 1. 
Note that the singlet superfield $\bar{E}$ is 
included to cancel $U(1)_m$ anomaly and that vector-like 
superfields $f \equiv (q,l), \bar{f} 
\equiv (\bar{q},\bar{l})$ are involved. 
Here we introduce the parameter $r$ which specifies 
the $U(1)_{R}$-charges for each superfields and 
we assume that the sum of R-charge of $f$ and $\bar{f}$ is 0. 
Under the standard model gauge group 
$G_{{\rm standard}} \equiv SU(3)_C 
\times SU(2)_L \times U(1)_Y$, 
$Q, \bar{U}, \bar{D}, L$ and $\bar{E}$ 
are neutral and $q, \bar{q}, l$ and $\bar{l}$ 
are transformed as 
\begin{equation}
q: ({\bf 3}, {\bf 1}, -2/3), \quad 
\bar{q}: ({\bf \bar{3}}, {\bf 1}, 2/3), \quad 
l: ({\bf 1}, {\bf 2}, 1), \quad 
\bar{l}: ({\bf 1}, {\bf 2}, -1), 
\end{equation}
respectively. 


\vspace{5mm}
\begin{center}
\framebox[3cm]{\large \bf Table 1}
\end{center}
\vspace{5mm}


The tree level superpotential consistent 
with the symmetries is 
\begin{equation}
W_{{\rm tree}} = \lambda_1 Q \bar{D} L + 
\frac{\kappa'}{M^2} (Q \bar{D} L) f\bar{f} + 
\frac{\lambda_2'}{M^5} 
(Q \bar{U} L) \bar{E} ({\rm det}Q\bar{Q}), 
\end{equation}
where $\bar{Q} = (\bar{U},\bar{D})$. 
In Eq. (2.2) the first term is a renormalizable term 
which has been treated in Ref. \cite{ADS85}. 
The second and the third terms are 
non-renormalizable terms which will be at most 
cubic terms of the $SU(3) \times SU(2)$ gauge invariant 
operators\footnote{Other non-renormalizable 
terms are highly suppressed by the scale $M$, 
which will be set to be the Planck scale later.}. 
The coupling constants $\lambda_2'$ 
and $\kappa'$ are taken to be of order $O(1)$. 
$\lambda_1$ is assumed to be very small 
so that the theory becomes weakly coupled.

To analyze the model we first consider the case 
in which the superpotential is absent. 
Under the condition that $g_3 \gg g_2 \gg g_1$, 
where $g_3, g_2$ and $g_1$ represent 
$SU(3), SU(2)$ and $U(1)_m$ gauge coupling constants, 
respectively, there exist the D-flat 
directions for $SU(3)$ and $SU(2)$ \cite{ADS85}
\begin{equation}
\langle Q \rangle = 
\left(
\begin{array}{cc}
a & 0 \\
0 & b \\
0 & 0 
\end{array}
\right), 
\langle \bar{U} \rangle = 
\left(
\begin{array}{c}
a \\
0 \\
0 
\end{array}
\right), 
\langle \bar{D} \rangle = 
\left(
\begin{array}{c}
0 \\
b \\
0 
\end{array}
\right), 
\langle L \rangle = (0, \sqrt{a^2-b^2}). 
\end{equation}
Here $a$ and $b$ are taken as real and positive 
parameters with $a \ge b$. 
Gauge symmetries are completely broken 
along these flat directions and 
the one-instanton effect induces the non-perturbative 
superpotential \cite{ADS85} 
\begin{equation} 
W_{{\rm dyn}} = \frac{\Lambda^7_3}{{\rm det}Q\bar{Q}}, 
\end{equation} 
where $\Lambda_3$ is the scale where $SU(3)$ 
gauge coupling blows up. 
Note that we consider the 
case $\Lambda_3 \gg \Lambda_2$, 
where $\Lambda_2$ is the scale where $SU(2)$ 
gauge coupling diverges 
\footnote{If we consider the 
case $\Lambda_2 \gg \Lambda_3$, 
supersymmetry is broken due to 
the quantum deformation of 
the moduli space \cite{IT}.}.

If we turn on the tree level superpotential, 
the flat directions are lifted. 
In our case VEVs of the fields are close to 
those of Ref. \cite{ADS85}, 
$v \sim \Lambda_3/\lambda_1^{1/7} \gg \Lambda_3$. 
The vacuum energy is $V \sim \lambda_1^2 v^4 > 0$ 
and then supersymmetry is broken. 
The moduli space is described in terms of 
the $SU(3) \times SU(2)$ gauge 
invariant operators \cite{LT} listed in Table 2. 
 From Eqs. (2.2) and (2.4) below 
the scale $\Lambda_3$ we have the effective 
superpotential 
\begin{eqnarray}
W_{{\rm eff}} &=& \lambda_1 X_1 
+ \frac{\Lambda^7_3}{X_3} 
+ \frac{\kappa'}{M^2} X_1 f\bar{f} 
+ \frac{\lambda_2'}{M^5} X_2 \bar{E} X_3, \\
&=& \lambda_1 v^2 Y + \frac{\lambda_1 v^4}{X} + 
\kappa Y f\bar{f} + \lambda_2 PNX, 
\end{eqnarray} 
where in the second equality we rescale 
the gauge invariant operators 
as $X_1 = v^2 Y, X_2 = v^2 N, X_3 = v^3 X$ 
and $\bar{E} = P$. 
Here we introduce the notations 
as $\kappa \equiv \kappa' (v/M)^2$ 
and $\lambda_2 \equiv \lambda_2' (v/M)^5$. 
Equation (2.6) represents the effective theory of 
the supersymmetry breaking sector. 
Since $N$ and $P$ have $U(1)_m$ charges $-1$ and $+1$, 
respectively, these superfields 
correspond to $\phi^-$ and $\phi^+$ 
in the minimal model \cite{DN}. 
Contrary to Ref. \cite{DN} we add no superfields to 
the effective superpotential. 
In the present model we have $\langle F_Y \rangle 
\sim \lambda_1 v^2$. Then, the supersymmetry breaking is 
communicated to the messenger fields $f$ and $\bar{f}$ 
through $\kappa Y f\bar{f}$ term. It is worth noting that 
the $U(1)_m$ gauge interaction and $\lambda_2 PNX$ term 
do not play an essential role in communicating the 
supersymmetry breaking to the messenger fields. 


\vspace{5mm} 
\begin{center} 
\framebox[3cm]{\large \bf Table 2} 
\end{center} 
\vspace{5mm} 


In order to analyze the scalar potential in our model, 
it is necessary to calculate the effective K\"ahler potential. 
Under the condition $\lambda_1 \ll 1$, we can calculate the effective 
K\"ahler potential using the procedure given by 
Poppitz and Randall \cite{PR} 
because the theory is weakly coupled and 
the gauge symmetries are completely broken. 
The effective K\"ahler potential is given by 
\begin{equation}
K = 3 \left( t + \frac{B}{t} \right), 
\end{equation}
where 
\begin{eqnarray}
t &\equiv& (A+\sqrt{A^2-B^3})^{1/3} 
+ (A-\sqrt{A^2-B^3})^{1/3}, \nonumber \\ 
A &\equiv& \frac{1}{2} ( X_1^{\dag}X_1+X_2^{\dag}X_2 ) 
= \frac{1}{2} v^4 (Y^{\dag} Y + N^{\dag} N ), \\ 
B &\equiv& \frac{1}{3} (X_3^{\dag}X_3)^{1/2} 
= \frac{1}{3} v^3 (X^{\dag} X)^{1/2}. \nonumber
\end{eqnarray}
The inverse of the effective K\"ahler metric is 
\begin{equation}
K^{j^* i} = 
\left(
\begin{array}{ccc}
\left( \frac{t}{v^2} \right)^2 + \frac{2}{t} Y^{\dag} Y 
& \frac{2}{t} Y^{\dag} N & \frac{2}{t} Y^{\dag} X \\
\frac{2}{t} N^{\dag} Y & \left( \frac{t}{v^2} \right)^2 + 
\frac{2}{t} N^{\dag} N & \frac{2}{t} N^{\dag} X \\
\frac{2}{t} X^{\dag} Y & \frac{2}{t} X^{\dag} N & 
\frac{2t}{v^3} |X| + \frac{2}{t} X^{\dag} X 
\end{array}
\right) 
\end{equation}
for $i, j = Y, N$ and $X$. 
Since $P ( = \bar{E}), f$ and $\bar{f}$ have no $SU(3)$ charge, 
their components of the effective K\"ahler potential are 
assumed to be of canonical form. From Eqs. (2.6) and (2.9) 
the scalar potential of the 
effective theory is given by
\begin{eqnarray}
V &=& W_{j^*} K^{j^*i} W_i + \frac{g^2_1}{2}(|P|^2-|N|^2)^2 
+ (M_P^2 |P|^2 + M_N^2 |N|^2), \nonumber \\ 
&=& \frac{2}{t} \left| \lambda_1 v^2 Y + 
\kappa Y f\bar{f} + 
2 \lambda_2 PNX -\frac{\lambda_1 v^4}{X} 
\right|^2 \nonumber \\
&& + \left( \frac{t}{v^2} \right)^2 
\left( \lambda_2^2 |X|^2|P|^2 
+ | \lambda_1 v^2 + \kappa f\bar{f} |^2 \right) 
+ \frac{2t}{v^3}|X| \left| \lambda_2 PN - 
\frac{\lambda_1 v^4}{X^2}  \right|^2 \nonumber \\ 
&& + \lambda_2^2 |N|^2|X|^2 
+ \kappa^2 |Y|^2 (|f|^2+|\bar{f}|^2) \nonumber \\ 
&& + \frac{g^2_1}{2}(|P|^2-|N|^2)^2 
+ (M_P^2 |P|^2 + M_N^2 |N|^2), 
\end{eqnarray}
where the second last term is $U(1)_m$ D-term, 
because $U(1)_m$ is gauged and $U(1)_m$ 
D-flatness condition is not imposed. 
The last term represents two-loop generated 
soft supersymmetry breaking mass term. 
Note that $M_P^2$ and $M_N^2$ are negative and of 
order $O\left( \left( \frac{g_1^2}{16\pi^2} \right)^2 
\lambda_1^2 v^2 \right)$.

By minimizing the scalar potential (2.10) 
under the conditions 
\begin{equation}
\kappa \gg \lambda_1, \quad  \frac{g^2_1}{16 \pi^2} 
\lambda_1 \gg \lambda_2, 
\end{equation}
we obtain VEVs 
which are in the vicinity of the $SU(3), SU(2)$ 
D-flat direction \cite{ADS85}, namely
\begin{eqnarray}
&&X = v_X + x, \quad |x| \ll v_X, \nonumber \\
&&Y = v_Y + y, \quad |y| \ll v_Y, \\
&&|f|, |\bar{f}|, |P| {\rm and} |N| \ll v, \nonumber
\end{eqnarray}
where $v_X \equiv a^2b^2/v^3, 
v_Y \equiv a^2 \sqrt{a^2-b^2} /v^2$ 
and $x, y$ represent the fluctuation 
around $v_X, v_Y$, respectively. 
In the minimization it is important that the effective 
K\"ahler potential has of the non-canonical form. 
We can easily derive 
\begin{equation}
\langle f \rangle = \langle \bar{f} \rangle = 0 
\end{equation}
 from the stationary conditions for 
$f$ and $\bar{f}$. 
This shows that QCD color is not broken. 
By calculating the minimization conditions 
with respect to $N$ and $P$ we obtain 
\begin{eqnarray}
|\langle P \rangle| &\simeq& \frac{1}{g_1} \sqrt{-M_P^2} \sim 
\frac{g_1}{16 \pi^2} \lambda_1 v, \\
|\langle N \rangle| &\simeq& \frac{\lambda_2}{\lambda_1} 
|\langle P \rangle| \sim \frac{g_1}{16 \pi^2} \lambda_2 v 
\end{eqnarray}
The order of VEVs of the fluctuation $x$ and $y$ is 
\begin{equation}
|\langle x \rangle| \sim |\langle y \rangle| \sim 
O \left( \left( \frac{g_1^2}{16 \pi^2} \right)^2 
\lambda_2^2 v \right) \ll v_X,v_Y. 
\end{equation}
As a consequence, our analysis is found to be 
self-consistent.

\section{Visible sector and Estimation of the scales}
\indent

Supersymmetry breaking in the messenger sector is transmitted 
to gauginos, squarks and sleptons in the visible sector 
radiatively through the interaction $\kappa Y f\bar{f}$ 
in Eq.(2.6). 
At one-loop we can obtain the masses for 
$SU(3)_C, SU(2)_L$ and $U(1)_Y$ gauginos 
\begin{equation}
m_{\lambda_i} \sim \frac{g_i^2}{16 \pi^2} 
\frac{\langle F_Y \rangle}{\langle Y \rangle} 
\sim \frac{g_i^2}{16 \pi^2} \lambda_1 v, 
\end{equation}
where $g_i$ stands for the corresponding gauge 
coupling of the standard model. 
Taking $m_{\lambda_i} = 10^{2.0 \pm 0.5}$GeV, we obtain 
\begin{equation}
\lambda_1 \left( \frac{v}{M_{{\rm Planck}}} \right) 
\sim 10^{-13.3 \pm 0.5}, 
\end{equation}
where $\frac{g_i^2}{16 \pi^2} \simeq 10^{-2.5}$ is used. 
On the other hand, the soft supersymmetry breaking 
masses for squarks and 
sleptons are induced at two-loop. 
They are given by 
\begin{equation}
m_{\phi_i}^2 \sim \left( \frac{g_i^2}{16 \pi^2} 
\frac{\langle F_Y \rangle}{\langle Y \rangle} \right)^2 
\sim \left( \frac{g_i^2}{16 \pi^2} \lambda_1 v \right)^2. 
\end{equation}

Here we focus on the estimation of the various 
scales. In the supersymmetry breaking sector we have a scale $M$ 
which supresses the non-renormalizable interactions in Eq.(2.2). 
It is natural to take this scale $M$ as $M_{{\rm Planck}}$. 
This implies that we have to take the effects of 
the gravitational interaction into account in calculating 
the masses for the gauginos, the squarks, and the sleptons. 
The scalar mass terms which are induced by gravity come from 
the D-term 
\begin{equation}
\int d^4 \theta \left( \frac{Q^{\dag}Q}{M_{{\rm Planck}}^2} 
+ \cdots \right) \Phi^{\dag}_i \Phi_i, 
\end{equation}
where $\Phi_i$ are superfields of the standard model 
and $i$ denotes flavor index. Therefore, the gravity-induced 
scalar masses are 
\begin{eqnarray}
m_{\phi_i}({\rm grav}) &\sim& 
\frac{\langle F \rangle}{M_{{\rm Planck}}}, \\
&\sim& \lambda_1 \left( \frac{v}{M_{{\rm Planck}}} \right)^2 
M_{{\rm Planck}} \ll 10^{2.5 \pm 0.5} {\rm GeV}, 
\end{eqnarray}
where $\langle F \rangle$ is the F-term in the supersymmetry 
breaking sector. The last inequality implies 
that we consider only the case in which the gravitational 
effects is negligible. From this inequality we obtain 
\begin{equation}
\lambda_1 \left( \frac{v}{M_{{\rm Planck}}} \right)^2 
\ll 10^{-15.8 \pm 0.5}, 
\end{equation}
where $M_{{\rm Planck}} \simeq 10^{18.3} {\rm GeV}$ is used.

On the other hand, gaugino mass terms which are induced 
by gravity arise via the term 
\begin{equation}
\int d^2 \theta \frac{X_1}{M_{{\rm Planck}}^3} 
W^{\alpha} W_{\alpha} = \int d^2 \theta 
\left( \frac{v}{M_{{\rm Planck}}} \right)^2 
\frac{Y}{M_{{\rm Planck}}} W^{\alpha} W_{\alpha}, 
\end{equation}
where $W^{\alpha}$ is a field strength superfield. 
Thus, the gravity-induced gaugino massess become 
\begin{equation}
m_{\lambda_i}({\rm grav}) \sim 
\lambda_1 \left( \frac{v}{M_{{\rm Planck}}} \right)^4 
M_{{\rm Planck}}. 
\end{equation}
 From Eq. (3.2) and the inequality (3.7) we find that 
\begin{equation}
m_{\lambda_i} \gg m_{\lambda_i}({\rm grav}). 
\end{equation}
Namely, the gauge-mediated contribution to the gaugino 
mass is dominant compared with the gravity-mediated 
contribution.

Taking the condition $\lambda_1 \ll \kappa \sim 
\left( \frac{v}{M_{{\rm Planck}}} \right)^2$ into account 
together with Eqs. (3.2) and (3.7), 
we obtain the allowed range of parameters 
\begin{eqnarray}
&& 10^{-4.4 \pm 0.2} \ll \frac{v}{M_{{\rm Planck}}} 
\ll 10^{-2.5}, \\
&& 10^{-10.8 \pm 0.5} \ll \lambda_1 
\ll 10^{-8.9 \pm 0.3}. 
\end{eqnarray} 
If we take $\lambda_1 \sim 10^{-9.3}$ as an example, 
various scales in the model are determined as,
\begin{eqnarray}
v &\sim& 10^{14.3} {\rm GeV}, \nonumber \\ 
\Lambda &\sim& 10^{13.0} {\rm GeV}, \nonumber \\ 
\sqrt{F} &\sim& \sqrt{\lambda_1 v^2} 
\sim 10^{9.6} {\rm GeV}, \\ 
m &\sim& \kappa v \sim 10^{6.3} {\rm GeV}, \nonumber \\ 
m_{\lambda_i} &\sim& 10^{2.5} {\rm GeV}, \nonumber 
\end{eqnarray}
where $m$ means the mass of $f$ and $\bar{f}$. 
Equation (3.5) also represents the order of the gravitino mass. 
In the present example the gravitino mass is $\sim 10^{1.5}$ GeV. 
We note that supersymmetry breaking scale 
$\sqrt{F}$ turns out to be the intermediate scale 
between the GUT scale and the soft 
supersymmetry breaking scale.

\section{Summary}
\indent

We showed that the messenger sector can be considered as 
the effective theory of supersymmetry breaking sector. 
No matter superfields and interactions are added 
in the messenger sector. In other words, 
all interactions in our messenger sector are derived from 
the supersymmetry breaking sector. Using this effective theory 
we also showed that supersymmetry breaking can be communicated 
to the visible sector without breaking QCD color.

In the present framework, the essential role of 
commuinicating the supersymmetry breaking to 
the visible sector is played by 
the $\kappa Y f\bar{f}$ term in the 
effective superpotential. We do not need to rely on 
the $U(1)_m$ gauge interaction and also on the $PNX$ 
term in the superpotential. This situation is in sharp 
contrast to that in Ref. \cite{DN}. In fact, 
when we do not introduce the field $P (= \bar{E})$ 
and the $U(1)_m$ gauge symmetry, we can make the model 
simpler. Even in that case the main point of 
our results remains unchanged.

There appear various scales in the present model. 
We also estimate the gravitational effects in 
calculating masses for gauginos, squarks, and sleptons. 
It is found that gauge-mediated contributions are 
dominant compared with gravity-mediated contributions. 
As a typical example, we found that gauge symmetry breaking 
scale ($v$) in 
supersymmetry breaking sector is $10^{14.3}$ GeV, 
the scale of $SU(3)$ dynamics ($\Lambda_3$) 
in $SU(3) \times SU(2)$ model 
is $10^{13.0}$GeV, supersymmetry breaking scale ($\sqrt{F}$) 
is $10^{9.6}$ GeV and the mass of messenger 
fields ($\kappa v$) is $10^{6.3}$ GeV.

In the visible sector of our model, as discussed in 
Ref. \cite{DN}, the masses for gauginos are induced at one-loop 
through the standard model gauge interaction. On the other hand, 
the masses for squarks, sleptons are induced at two-loop 
through the standard model gauge interaction. 
Therefore, FCNC are naturally suppressed because these masses 
are proportional to the square of the flavor-blind standard 
model gauge coupling constants.

The present model will provide a useful guide for constructing 
the phenomenologically viable models of 
the gauge-mediated supersymmetry breaking. 


\begin{flushleft}
{\bf Acknowledgements}
\end{flushleft}

One of the authors, N.H. would like to 
thank Prof. T. Yanagida and Dr. K.-I. Izawa for 
useful discussions. 
We thank Prof. S. Kitakado for 
careful reading of the manuscript.


\newpage


\begin{center}
{\LARGE Figure Captions}
\end{center}
\begin{flushleft}
{\bf Table 1}
\end{flushleft}
The representations and charges of matter fields in 
supersymmetry breaking sector. The parameter $r$, 
which specifies $U(1)_R$-charges of each superfield, 
is left arbitrary. 
$r_f$ and $r_{\bar{f}}$ are R-charges of $f$ and $\bar{f}$, 
respectively. 
\vspace{5mm}
\begin{flushleft}
{\bf Table 2}
\end{flushleft}
The gauge invariant operators which describe 
the moduli space.

\newpage


\begin{center}
{\bf Table 1}
\end{center}
\begin{center}
    \begin{tabular}{|c|c|c|c|}
\hline
Particle & $SU(3) \times SU(2)$ & $U(1)_m$ & $U(1)_R$ \\
\hline
$Q$ & ( 3 , 2 ) & 1/6 & $r$ \\

$\bar{U}$ & ( $\bar{3}$ , 1 ) & $-2/3$ & $-4-4r$ \\

$\bar{D}$ & ( $\bar{3}$ , 1 ) & 1/3 & $2+2r$ \\

$L$ & ( 1 , 2 ) & $-1/2$ & $-3r$ \\

$\bar{E}$ & ( 1 , 1 ) & 1 & $8+6r$ \\
\hline
$f$ & ( 1 , 1 ) & 0 & $r_f$ \\
$\bar{f}$ & ( 1 , 1 ) & 0 & $r_{\bar{f}}$ \\
\hline
\end{tabular}
\end{center}

\vspace{1cm}

\begin{center}
{\bf Table 2}
\end{center}
\begin{center}
    \begin{tabular}{|c|c|c|}
\hline
 & $U(1)_m$ & $U(1)_R$ \\
\hline 
\vphantom{\Big(}
$X_1 = Q \bar{D} L$ & 0 & 2 \\
$X_2 = Q \bar{U} L$ & $-1$ & $-4-6r$ \\
$X_3 = {\rm det} Q \bar{Q}$ & 0 & $-2$ \\
$\bar{E}$ & 1 & $8+6r$ \\
\hline 
\vphantom{\Big(}
$f \bar{f}$ & 0 & 2 \\
\hline
\end{tabular}
\end{center}


\begin{thebibliography}{99}
\bibitem{Witten82}E. Witten, Nucl. Phys. {\bf B212} 
(1982) 253. 
\bibitem{Nilles}H.P. Nilles, Phys. Rep. {\bf 110} 
(1984) 1. 
\bibitem{GMSB}M. Dine, W. Fischer, and M. Srednicki, 
Nucl. Phys. {\bf B189} (1981) 575. \\
S. Dimopoulos and S. Raby, Nucl. Phys. {\bf B192} (1981) 353. \\
M. Dine and W. Fischer, Phys. Lett. {\bf B110} (1982) 227. \\
M. Dine and M. Srednicki, Nucl. Phys. {\bf B202} (1982) 238. \\
M. Dine and W. Fischer, Nucl. Phys. {\bf B204} (1982) 346. \\
L. Alvarez-Gaume, M. Claudson, and M. Wise, Nucl. Phys. 
{\bf B207} (1982) 96. \\
C.R. Nappi and B.A. Orvut, Phys. Lett. {\bf B113} (1982) 175. \\
S. Dimopoulos and S. Raby, Nucl. Phys. {\bf B219} (1983) 479. 
\bibitem{DN}M. Dine and A.E. Nelson, Phys. Rev. 
{\bf D48} (1993) 1277, hep-ph/9303230. \\
M. Dine, A.E. Nelson, and Y. Shirman, Phys. Rev. 
{\bf D51} (1995) 1362, hep-ph/9408384. \\
M. Dine, A.E. Nelson, Y. Nir, and Y. Shirman, Phys. Rev. 
{\bf D53} (1996) 2658, hep-ph/9507378. 
\bibitem{Randall}I. Dasgupta, B.A. Dobrescu, and L. Randall, 
hep-ph/9607487. 
\bibitem{Many}M. Dine, Y. Nir, and Y. Shirman, hep-ph/9607397. \\
S.P. Martin, hep-ph/9608224. \\
S. Dimopoulos and G.F. Giudice, hep-ph/9609344. \\
S. Dimopoulos, S. Thomas, and J.D. Wells, hep-ph/9609434. 
\bibitem{ADS85}I. Affleck, M. Dine and N. Seiberg, 
Nucl. Phys. {\bf B256} (1985) 557. 
\bibitem{IT}K. Intriligator and S. Thomas, Nucl. Phys. 
{\bf B473} (1996) 121, hep-th/9603158. 
\bibitem{LT}M.A. Luty and W. Taylor, Phys. Rev. 
{\bf D53} (1996) 3399, hep-th/9506098. 
\bibitem{PR}E. Poppitz and L. Randall, Phys. Lett. 
{\bf B336} (1994) 402, hep-th/9407185.
\end{thebibliography}
\end{document}